\documentstyle[aps,12pt]{revtex}

\widetext
\draft
\tighten
\preprint{EFUAZ FT-97-44}

\begin{document}

\title{Self/Anti-Self Charge Conjugate
States for $j=1/2$ and $j=1$\thanks{Presented at the XXXIX Congreso
Nacional de F\'{\i}sica, Oaxaca, Oax., M\'exico, Oct.  14-18, 1996, the
Primeras Jornadas de Investigaci\'on, Zacatecas, Nov.  18-19, 1996 and the
V Reunion Anual de DGFM de la Sociedad Mexicana de F\'{\i}sica, M\'exico,
Apr. 17-18, 1997}}

\author{{\bf Valeri V. Dvoeglazov}}

\address{Escuela de F\'{\i}sica, Universidad Aut\'onoma de Zacatecas\\
Apartado Postal C-580, Zacatecas 98068 Zac., M\'exico\\
Internet address: VALERI@CANTERA.REDUAZ.MX\\
URL: http://cantera.reduaz.mx/\~\,valeri}

\date{April 18, 1997}

\maketitle

\begin{abstract}
We briefly review recent achievements in the theory of neutral particles
(the Majorana-McLennan-Case-Ahluwalia construct for self/anti-self
charge conjugate states for $j=1/2$ and $j=1$ cases).  Among new results we
present a theoretical construct in which a fermion and an antifermion
have the same intrinsic parity; discuss phase transformations
and find relations between the Majorana-like field operator
$\nu$, given by Ahluwalia, and the Dirac field operator. Also we
give explicit forms of the $j=1$ ``spinors" in the Majorana
representation.
\end{abstract}


\newpage

\large{
The construct for self/anti-self charge conjugate states defined in the
momentum representation has been proposed in refs.~\cite{DVA,DVO}.
This is the straightforward development of the Majorana ideas~\cite{MAJ}
and the ideas of McLennan~\cite{MCL} and Case~\cite{Case}.

Let us present the previous results:

\begin{itemize}

\item
Self/anti-self charge conjugate spinors have been defined in the
$(1/2,0)\oplus (0,1/2)$ representation in the momentum space~[1c]:
\begin{eqnarray}
\lambda^{S,A} = \pmatrix{\pm i\Theta \phi_L^\ast (p^\mu)\cr
\phi_L (p^\mu)\cr}\quad,\quad
\rho^{S,A} = \pmatrix{\phi_R (p^\mu)\cr \mp i\Theta \phi_R^\ast (p^\mu)\cr}
\quad
\end{eqnarray}
and have been named as the type-II spinors.
They are eigenstates of the charge conjugation operator:
\begin{eqnarray}
S^c_{[1/2]} = e^{i\theta_c} \pmatrix{0 & i\Theta\cr
-i\Theta & 0\cr} {\cal K}\quad,\quad \Theta \equiv -i\sigma_2 =
\pmatrix{0&-1\cr
1&0}\quad;
\end{eqnarray}
\begin{mathletters}
\begin{eqnarray}
S_{[1/2]}^c \lambda^{S,A} (p^\mu) &=& \pm \lambda^{S,A} (p^\mu)\quad,\\
S_{[1/2]}^c \rho^{S,A} (p^\mu) &=& \pm \rho^{S,A} (p^\mu)\quad.
\end{eqnarray}
\end{mathletters}
Similar states  (to a certain extent) can be constructed in the higher
representations of the Lorentz group, e.g., in the $(j,0)\oplus (0,j)$
representation, $j>1/2$.

\item
The field operator
\begin{eqnarray}
\nu^{DL} (x^\mu) &=& \sum_\eta \int \frac{d^3 {\bf p}}{(2\pi)^3}
{1\over 2E_p} \left [ \lambda^S_\eta (p^\mu) a_\eta (p^\mu) \exp (-ip\cdot
x) \right.\nonumber\\
&+&\left.\lambda^A_\eta (p^\mu) b_\eta^\dagger (p^\mu) \exp
(+ip\cdot x)\right ]\quad
\end{eqnarray}
has been proposed for this sort of states~[1c].

\item
$\lambda$ and $\rho$ spinors are {\it not} eigenspinors of the
$(j,0)\oplus (0,j)$ helicity operator
\begin{equation}
h= \pmatrix{{\bf J}\cdot \hat {\bf n} &0\cr
0& {\bf J}\cdot \hat {\bf n}\cr}
\end{equation}
(by the definition, indeed, because  $\Theta_{[j]}
{\bf J} \Theta^{-1}_{[j]} = -{\bf J}^\ast$). The new quantum
number ({\it chiral helicity}) corresponding to the operator
$\eta=-\Gamma^5 h$ has been introduced.

\item
$\lambda$ and $\rho$ spinors are {\it not} eigenspinors of the parity
operator, see formulas (36a,b) in ref.~[1c]. ``This is not related to the
fact that $S_{[1/2]}^c$ and $S_{[1/2]}^s$ do not commute. Since
$S_{[1/2]}^c$ is {\it not} linear, it is possible to have a simultaneous
set of eigenspinors, but such a set does not have its eigenspinors of
type-II",  in the opinion of D. V. Ahluwalia, ref.~[1c].

\item
The introduction of the interaction in an usual manner
(``covariantization" $\partial_\mu \rightarrow \nabla_\mu = \partial_\mu
-ieA_\mu$) was found to be impossible for these states because phase
transformations which correspond to this ``covariantization" would lead
to the consequence that the spinors would {\it not} keep their property to
be self/anti-self charge conjugate spinors.

\item
Simple dynamical equations for $\lambda$ and $\rho$ spinors have been
obtained~[2d] on the basis of a new form of the Ryder-Burgard relation
(which connects the left- and right- parts of the bispinors in the frame
with zero momentum~\cite{Ryder,DVA1}). Here they are:
\begin{mathletters}
\begin{eqnarray}
i\gamma^\mu \partial_\mu \lambda^S (x) - m\rho^A (x) &=& 0\quad,\\
i\gamma^\mu \partial_\mu \rho^A (x) - m\lambda^S (x) &=& 0\quad,\\
i\gamma^\mu \partial_\mu \lambda^A (x) + m\rho^S (x) &=& 0\quad,\\
i\gamma^\mu \partial_\mu \rho^S (x) + m\lambda^A (x) &=& 0\quad.
\end{eqnarray}
\end{mathletters}
In fact they can be written in eight-component form, see
also the old works~\cite{Markov,Belin} and the recent
works~\cite{Ziino,Robson}.

\item
The connection with the Dirac spinors has been found~[2a,b]. For instance,
\begin{eqnarray}
\pmatrix{\lambda^S_\uparrow (p^\mu) \cr \lambda^S_\downarrow (p^\mu) \cr
\lambda^A_\uparrow (p^\mu) \cr \lambda^A_\downarrow (p^\mu)\cr} = {1\over
2} \pmatrix{1 & i & -1 & i\cr -i & 1 & -i & -1\cr 1 & -i & -1 & -i\cr i&
1& i& -1\cr} \pmatrix{u_{+1/2} (p^\mu) \cr u_{-1/2} (p^\mu) \cr
v_{+1/2} (p^\mu) \cr v_{-1/2} (p^\mu)\cr}\quad.\label{connect}
\end{eqnarray}
See also ref.~\cite{Ziino}.

\item
The sets of $\lambda$ spinors and of $\rho$ spinors are claimed~[1c] to be
{\it bi-orthonormal} sets each in the mathematical sense,  provided
that overall phase factors of 2-spinors $\theta_1 +\theta_2 = 0$ or $\pi$.
For instance, on the classical level $\bar \lambda^S_\uparrow
\lambda^S_\downarrow = 2iN^2 \cos ( \theta_1 + \theta_2 )$.
Corresponding commutation relations for this
type of states have also been proposed.

\item
The Lagrangian for $\lambda$ and $\rho$-type  $j=1/2$ states was
given~[2d,formula(24)].

\item
While in the massive case there are four $\lambda$-type spinors, two
$\lambda^S$ and two $\lambda^A$ (the $\rho$ spinors are connected by
certain relations with the $\lambda$ spinors for any spin case),  in a
massless case $\lambda^S_\uparrow$ and $\lambda^A_\uparrow$ identically
vanish, provided that one takes into account that $\phi_L^{\pm 1/2}$ are
eigenspinors of ${\bbox \sigma}\cdot \hat {\bf n}$, the
$2\times 2$ helicity operator.

\item
It was noted the possibility of the generalization of the concept of the
Fock space, which leads to the ``doubling" Fock space~\cite{Ziino}.

\item
There does not exist the self/anti-self charge conjugate ``spinors" in the
$(1,0) \oplus (0,1)$ representation. Therefore, $\Gamma^5 S^c_{[1]}$
self/anti-self  conjugate objects have been defined there.

\item
The {\it commutator} of the operations $U^s_{[1/2]}$ and $U^c_{[1/2]}$ in
the Fock space may be equal to zero when acting on the Majorana states.
The parity operator of the Fock space is the function of the charge
operator~\cite{Nigam}.

\item
Several explicit constructs of the Bargmann-Wightman-Wigner-type
theories~\cite{BWW} have been presented in~\cite{DVA1,DVA,DVO,Ziino}.

\end{itemize}

We continue researches in the area of the physics of neutral particles
because the present-day standard models do not provide any adequate
formalism for describing neutrino and photon. Among new results we now
present:

\begin{itemize}

\item
It was shown that the covariant derivative (and, hence, the
interaction) can be introduced in this construct in the following way:
\begin{equation}
\partial_\mu \rightarrow \nabla_\mu = \partial_\mu - ig \L^5 A_\mu\quad,
\end{equation}
where $\L^5 = \mbox{diag} (\gamma^5 \quad -\gamma^5)$, the $8\times 8$
matrix. With respect to the transformations
\begin{mathletters}
\begin{eqnarray}
\lambda^\prime (x)
\rightarrow (\cos \alpha -i\gamma^5 \sin\alpha) \lambda
(x)\quad,\label{g10}\\
\overline \lambda^{\,\prime} (x) \rightarrow
\overline \lambda (x) (\cos \alpha - i\gamma^5
\sin\alpha)\quad,\label{g20}\\
\rho^\prime (x) \rightarrow  (\cos \alpha +
i\gamma^5 \sin\alpha) \rho (x) \quad,\label{g30}\\
\overline \rho^{\,\prime} (x) \rightarrow  \overline \rho (x)
(\cos \alpha + i\gamma^5 \sin\alpha)\quad\label{g40}
\end{eqnarray}
\end{mathletters}
the spinors retain their properties to be self/anti-self charge conjugate
spinors and the proposed Lagrangian~[2d, p. 1472] remains to be invariant.
This tells us that while self/anti-self charge conjugate states has
zero eigenvalues of the ordinary (scalar) charge operator but they can
possess the axial charge (cf.  with the discussion of~\cite{Ziino} and
the old idea of R. E. Marshak).

In fact, from this consideration one can recover the Feynman-Gell-Mann
equation (and its charge-conjugate equation). It is re-written in the
two-component form
\begin{eqnarray} \cases{\left [\pi_\mu^- \pi^{\mu\,-}
-m^2 -{g\over 2} \sigma^{\mu\nu} F_{\mu\nu} \right ] \chi (x)=0\quad, &\cr
\left [\pi_\mu^+ \pi^{\mu\,+} -m^2
+{g\over 2} \widetilde\sigma^{\mu\nu} F_{\mu\nu} \right ] \phi (x)
=0\quad, &\cr}\label{iii}
\end{eqnarray}
where already one has $\pi_\mu^\pm =
i\partial_\mu \pm gA_\mu$, \, $\sigma^{0i} = -\widetilde\sigma^{0i} =
i\sigma^i$, $\sigma^{ij} = \widetilde\sigma^{ij} = \epsilon_{ijk}
\sigma^k$ and $\nu^{^{DL}} (x) =\mbox{column} (\chi \quad \phi )$.

\item
Next, because the transformations
\begin{mathletters}
\begin{eqnarray}
\lambda_S^\prime (p^\mu) &=& \pmatrix{\Xi &0\cr 0&\Xi} \lambda_S (p^\mu)
\equiv \lambda_A^\ast (p^\mu)\quad,\quad\\
\lambda_S^{\prime\prime} (p^\mu) &=& \pmatrix{i\Xi &0\cr 0&-i\Xi} \lambda_S
(p^\mu) \equiv -i\lambda_S^\ast (p^\mu)\quad,\quad\\
\lambda_S^{\prime\prime\prime} (p^\mu) &=& \pmatrix{0& i\Xi\cr
i\Xi &0\cr} \lambda_S (p^\mu) \equiv i\gamma^0 \lambda_A^\ast
(p^\mu)\quad,\quad\\
\lambda_S^{IV} (p^\mu) &=& \pmatrix{0& \Xi\cr
-\Xi&0\cr} \lambda_S (p^\mu) \equiv \gamma^0\lambda_S^\ast
(p^\mu)\quad
\end{eqnarray}
\end{mathletters}
with the $2\times 2$ matrix $\Xi$ defined as ($\phi$ is the azimuthal
angle  related with ${\bf p} \rightarrow {\bf 0}$)
\begin{equation}
\Xi = \pmatrix{e^{i\phi} & 0\cr 0 &
e^{-i\phi}\cr}\quad,\quad \Xi \Lambda_{R,L} (\overcirc{p}^\mu \leftarrow
p^\mu) \Xi^{-1} = \Lambda_{R,L}^\ast (\overcirc{p}^\mu \leftarrow
p^\mu)\,\,\, ,
\end{equation}
and corresponding transformations for
$\lambda^A$ do {\it not} change the properties of bispinors to be in the
self/anti-self charge conjugate spaces, the Majorana-like field operator
($b^\dagger \equiv a^\dagger$) admits additional phase (and, in general,
normalization) transformations:
\begin{equation} \nu^{ML\,\,\prime}
(x^\mu) = \left [ c_0 + i({\bbox \tau}\cdot  {\bf c}) \right
]\nu^{ML\,\,\dagger} (x^\mu) \quad, \end{equation} where $c_\alpha$ are
arbitrary parameters. The ${\bbox \tau}$ matrices are defined over the
field of $2\times 2$ matrices\footnote{This concept is closely related
with the Wigner's concept of the {\it sign} spin, which was discussed
recently by M. Moshinsky~\cite{Mosh}. In general, this notation was used
extensively in the earlier works of many researchers.} and the Hermitian
conjugation operation is assumed to act on the $c$- numbers as the complex
conjugation. One can parametrize $c_0 = \cos\phi$ and ${\bf c} = {\bf n}
\sin\phi$ and, thus, define the $SU(2)$ group of phase transformations.
One can select the Lagrangian which is composed from the both field
operators (with $\lambda$ spinors and $\rho$ spinors)
and which remains to be
invariant with respect to this kind of transformations.  The conclusion
is: it is permitted a non-Abelian construct which is based on
the spinors of the Lorentz group only (cf. with the old ideas of T. W.
Kibble and R. Utiyama) .  This is not surprising because both the $SU(2)$
group and $U(1)$ group are  the sub-groups of the extended Poincar\'e group
(cf.~\cite{Ryder}).  Another non-Abelian model was proposed in the
$(1,0)\oplus (0,1)\oplus (1/2,1/2)$ by T.  Barrett (e.g.,
ref.~\cite{Barrett}) and, recently, in the $(1,0)\oplus (0,1)$
representation by J.-P.  Vigier {\it et al.}~\cite{Evans}.

\item
The new construct has been presented in which the fermion and its
antifermion may have the same intrinsic parities~[1c,2f].  We can deduce
the following properties of creation (annihilation) operators in the Fock
space:  \begin{mathletters} \begin{eqnarray} U^s_{[1/2]} a_\uparrow ({\bf
p}) (U^s_{[1/2]})^{-1} &=& - ia_\downarrow (-  {\bf p})\, ,\quad
U^s_{[1/2]} a_\downarrow ({\bf p}) (U^s_{[1/2]})^{-1} = + ia_\uparrow
(- {\bf p})\, ,\nonumber\\
&&\\
U^s_{[1/2]} b_\uparrow^\dagger ({\bf p}) (U^s_{[1/2]})^{-1} &=&
+ i b_\downarrow^\dagger (- {\bf p})\, ,\quad
U^s_{[1/2]} b_\downarrow^\dagger ({\bf p}) (U^s_{[1/2]})^{-1} =
- i b_\uparrow (- {\bf p})\, ,\nonumber\\
&&
\end{eqnarray} \end{mathletters}
what signifies that the states created by the operators $a^\dagger
({\bf p})$ and $b^\dagger ({\bf p})$ have very different properties
with respect to the space inversion operation, comparing to
Dirac states ($\pm$ stand for denoting the positive- (negative) energy
states). Namely,
\begin{mathletters} \begin{eqnarray} U^s_{[1/2]} \vert {\bf
p},\,\uparrow >^+ &=& + i \vert -{\bf p},\, \downarrow >^+\, ,\quad
U^s_{[1/2]} \vert {\bf p},\,\uparrow >^- = + i
\vert -{\bf p},\, \downarrow >^-\, ,\\
U^s_{[1/2]} \vert {\bf p},\,\downarrow >^+ &=& - i \vert -{\bf p},\,
\uparrow >^+\, ,\quad
U^s_{[1/2]} \vert {\bf p},\,\downarrow >^- =  - i
\vert -{\bf p},\, \uparrow >^-\, .
\end{eqnarray}
\end{mathletters}
For the charge conjugation operation in the Fock space we have
two physically different possibilities. The first one
\begin{mathletters}
\begin{eqnarray}
U^c_{[1/2]} a_\uparrow ({\bf p}) (U^c_{[1/2]})^{-1} &=& + b_\uparrow
({\bf p})\, ,\quad
U^c_{[1/2]} a_\downarrow ({\bf p}) (U^c_{[1/2]})^{-1} = + b_\downarrow
({\bf p})\, ,\\
U^c_{[1/2]} b_\uparrow^\dagger ({\bf p}) (U^c_{[1/2]})^{-1} &=&
-a_\uparrow^\dagger ({\bf p})\, ,\quad
U^c_{[1/2]} b_\downarrow^\dagger ({\bf p})
(U^c_{[1/2]})^{-1} = -a_\downarrow^\dagger ({\bf p})\,
\end{eqnarray}
\end{mathletters}
is, in fact,  reminiscent with the Dirac construct.
The action of this operator on the physical states are
\begin{mathletters}
\begin{eqnarray}
U^c_{[1/2]} \vert {\bf p}, \, \uparrow >^+ &=& + \,\vert {\bf p},\,
\uparrow >^- \quad,\quad
U^c_{[1/2]} \vert {\bf p}, \, \downarrow >^+ = + \, \vert {\bf p},\,
\downarrow >^- \quad,\quad\\
U^c_{[1/2]} \vert {\bf p}, \, \uparrow >^-
&=&  - \, \vert {\bf p},\, \uparrow >^+ \quad,\quad
U^c_{[1/2]} \vert
{\bf p}, \, \downarrow >^- = - \, \vert {\bf p},\, \downarrow >^+ \quad.
\end{eqnarray} \end{mathletters}
But, one can also build the charge conjugation operator in the
Fock space which acts, {\it e.g.}, in the following manner:
\begin{mathletters}
\begin{eqnarray}
\widetilde U^c_{[1/2]} a_\uparrow ({\bf p}) (\widetilde U^c_{[1/2]})^{-1}
&=& - b_\downarrow ({\bf p})\, ,\quad \widetilde U^c_{[1/2]}
a_\downarrow ({\bf p}) (\widetilde U^c_{[1/2]})^{-1} = - b_\uparrow
({\bf p})\, ,\\
\widetilde U^c_{[1/2]} b_\uparrow^\dagger ({\bf p})
(\widetilde U^c_{[1/2]})^{-1} &=& + a_\downarrow^\dagger ({\bf
p})\, ,\quad
\widetilde U^c_{[1/2]} b_\downarrow^\dagger ({\bf p})
(\widetilde U^c_{[1/2]})^{-1} = + a_\uparrow^\dagger ({\bf p})\, ,
\end{eqnarray}
\end{mathletters}
and, therefore,
\begin{mathletters}
\begin{eqnarray}
\widetilde U^c_{[1/2]} \vert {\bf p}, \, \uparrow >^+ &=& - \,\vert {\bf
p},\, \downarrow >^- \quad,\quad
\widetilde U^c_{[1/2]} \vert {\bf p}, \, \downarrow
>^+ = - \, \vert {\bf p},\, \uparrow >^- \quad,\quad\\
\widetilde U^c_{[1/2]} \vert
{\bf p}, \, \uparrow >^- &=& + \, \vert {\bf p},\, \downarrow >^+
\quad,\quad
\widetilde U^c_{[1/2]} \vert {\bf p}, \, \downarrow >^- = + \, \vert {\bf
p},\, \uparrow >^+ \quad.
\end{eqnarray}
\end{mathletters}
One can convince ourselves  by straightforward
verification  in the correctness of the
assertions made in~\cite{DVA} (see also the old paper~\cite{Nigam}) that
it is possible a situation when the operators of the space inversion and
the charge conjugation commute each other in the Fock space. For instance,
\begin{mathletters}
\begin{eqnarray}
U^c_{[1/2]} U^s_{[1/2]} \vert {\bf
p},\, \uparrow >^+ &=& + i U^c_{[1/2]}\vert -{\bf p},\, \downarrow >^+ =
+ i \vert -{\bf p},\, \downarrow >^- \quad,\\
U^s_{[1/2]} U^c_{[1/2]} \vert {\bf
p},\, \uparrow >^+ &=& U^s_{[1/2]}\vert {\bf p},\, \uparrow >^- = + i
\vert -{\bf p},\, \downarrow >^- \quad.
\end{eqnarray} \end{mathletters}
The second choice of the charge conjugation operator answers for the case
when the $\widetilde U^c_{[1/2]}$ and $U^s_{[1/2]}$ operations
anticommute:
\begin{mathletters} \begin{eqnarray}
\widetilde U^c_{[1/2]} U^s_{[1/2]} \vert {\bf p},\, \uparrow >^+ &=&
+ i \widetilde U^c_{[1/2]}\vert -{\bf
p},\, \downarrow >^+ = -i \, \vert -{\bf p},\, \uparrow >^- \quad,\\
U^s_{[1/2]} \widetilde U^c_{[1/2]} \vert {\bf p},\, \uparrow >^+ &=& -
U^s_{[1/2]}\vert {\bf p},\, \downarrow >^- = + i \, \vert -{\bf p},\,
\uparrow >^- \quad.
\end{eqnarray} \end{mathletters}
Next, one can compose states which would have somewhat similar
properties to those which we have become accustomed.
The states $\vert {\bf p}, \,\uparrow >^+ \pm
i\vert {\bf p},\, \downarrow >^+$ answer for positive (negative) parity,
respectively.  But, what is important, {\it the antiparticle states}
(moving backward in time) have the same properties with respect to the
operation of space inversion as the corresponding {\it particle states}
(as opposed to the $j=1/2$ Dirac particles).  This is again in accordance
with the analysis of Nigam and Foldy~\cite{Nigam}, and Ahluwalia~[1c].
The states which are eigenstates of the charge conjugation operator in the
Fock space are
\begin{equation} U^c_{[1/2]} \left ( \vert {\bf p},\,
\uparrow >^+ \pm i\, \vert {\bf p},\, \uparrow >^- \right ) = \mp i\,
\left ( \vert {\bf p},\, \uparrow >^+ \pm i\, \vert {\bf p},\, \uparrow
>^- \right ) \quad.  \end{equation} There is no a simultaneous set of
states which were ``eigenstates" of the operator of the space inversion
and of the charge conjugation $U^c_{[1/2]}$.

\item
We have found the Majorana representation of the Barut-Muzinich-Williams
matrices and the spinors of the (modified) Weinberg formulation (the
momentum-space functions in the $(1,0)\oplus (0,1)$ representation space).
In this representation all the matrices are the real matrices. The
matrix of the unitary transformation is:
\begin{mathletters}
\begin{eqnarray} U &=& {1\over 2\sqrt{2}}\pmatrix{(1-i) +(1+i) \Theta&
-(1-i) +(1+i) \Theta\cr (1+i) +(1-i)\Theta& -(1+i) +
(1-i)\Theta\cr}\quad,\\ U^\dagger &=& {1\over
2\sqrt{2}}\pmatrix{(1+i)+(1-i)\Theta& (1-i)+(1+i)\Theta\cr -(1+i)
+(1-i)\Theta& -(1-i) +(1+i) \Theta\cr}\quad.
\end{eqnarray}
\end{mathletters}
As a result we arrive, $\gamma_{\mu\nu}^{^{MR}}=U\gamma_{\mu\nu}^{^{CR}}
U^\dagger$:
\begin{mathletters} \begin{eqnarray}
\gamma_{00}^{^{MR}} &=& \pmatrix{0&\Theta\cr \Theta
&0\cr}\, ,\quad
\gamma_{01}^{^{MR}} = \gamma_{10}^{^{MR}} =\pmatrix{0&-J_1
\Theta\cr -J_1 \Theta&0\cr}\, ,\\
\gamma_{02}^{^{MR}} &=& \gamma_{20}^{^{MR}} =\pmatrix{iJ_2 \Theta& 0\cr
0 & -iJ_2 \Theta\cr}\, ,\quad
\gamma_{03}^{^{MR}} = \gamma_{30}^{^{MR}} =\pmatrix{0&-J_3
\Theta\cr -J_3 \Theta&0\cr}\, , \nonumber\\
&&\\
\gamma_{ij}^{^{MR}} &=& \gamma_{ji}^{^{MR}} = {1\over 2}
\pmatrix{i (J_{ij}^\ast -J_{ij} ) \Theta & (J_{ij}^\ast + J_{ij})\Theta\cr
(J_{ij}^\ast +J_{ij} )\Theta & -i (J_{ij}^\ast -J_{ij})
\Theta\cr}\, , \\
&& \mbox{and}\quad \gamma_5^{^{MR}} = \pmatrix{0&i\openone\cr
-i\openone & 0\cr}\, .
\end{eqnarray} \end{mathletters}
The $3\times 3$ matrix $\Theta$ corresponds to the Wigner operator in the
spin-1 representation
\begin{equation}
\theta = \pmatrix{0&0&1\cr
0&-1&0\cr
1&0&0\cr}\quad.
\end{equation}
If one writes
\begin{mathletters}
\begin{eqnarray}
u^{^{MR}} (p^\mu) &=& {1\over 2} \pmatrix{\phi_{_L} +
\Theta \phi_{_R}\cr \phi_{_L} +\Theta \phi_{_R}\cr} +{i\over 2}
\pmatrix{-\phi_{_L} +\Theta \phi_{_R}\cr \phi_{_L} -\Theta\phi_{_R}\cr} =
{\cal U}^+ +i{\cal V}^+\quad,\label{usp}\\
v^{^{MR}} (p^\mu) &=& {1\over
2} \pmatrix{-\phi_{_L} + \Theta \phi_{_R}\cr -\phi_{_L} +\Theta
\phi_{_R}\cr} +{i\over 2} \pmatrix{\phi_{_L} +\Theta \phi_{_R}\cr
-\phi_{_L} -\Theta\phi_{_R}\cr}= {\cal U}^- +i{\cal V}^-\quad. \label{vsp}
\end{eqnarray} \end{mathletters}
one can see that
\begin{equation}
v^{^{MR}} (p^\mu) = \gamma_5^{^{MR}}
u^{^{MR}} (p^\mu) = i\gamma_5^{^{WR}}\gamma_0^{^{WR}} u^{^{MR}} (p^\mu)
=\pmatrix{0&i\openone \cr -i\openone &0\cr} u^{^{MR}} (p^\mu)\quad.
\label{connect1}
\end{equation}
Surprisingly, we have
\begin{mathletters}
\begin{eqnarray}
{\cal U}_\uparrow^+ (p^\mu) &=& {\cal U}_\downarrow^+ (p^\mu)\, , \quad
{\cal V}_\uparrow^+ (p^\mu) = -{\cal V}_\downarrow^+ (p^\mu)\quad,\\
&&\mbox{but}\quad {\cal U}_\rightarrow^+ (p^\mu) = 0\quad,\quad {\cal
V}_\rightarrow^+ (p^\mu) \neq 0\, .
\end{eqnarray} \end{mathletters}
While the ``longitudinal" bispinor $u_\rightarrow (p^\mu)$ has only the
imaginary part in this reprersentation, the negative-energy bispinor
$v_\rightarrow (p^\mu)$ has only the real part.

Finally, it is interesting to note that the $\lambda^{S(A)} (p^\mu)$ and
$\rho^{S(A)} (p^\mu)$ spinors become the pure real (pure imaginary)
spinors in the momentum space representation for both $j=1/2$ and $j=1$
case.

\item
Furthermore, we have found some connections between the Dirac field
operator and the Majorana-like operator composed of $\lambda^{S,A}$
spinors. If one uses relations (\ref{connect}) between the self/anti-self
charge conjugate spinors and the Dirac spinors (together with the
identities between $\lambda$ and $\rho$ spinors) one can deduce:
\begin{equation}
\Psi^{Dirac} (x^\mu) = (1 +{ i\gamma^\mu \partial_\mu \over m} )
\nu^{^{ML}} (x^\mu) \quad.
\end{equation}
The commutation relations (for the creation/annihilation
operators of  self/anti-self charge conjugate states) may be slightly
different comparing to those presented in~[1c] but the set of the states
is still {\it bi-orthonormal}.

Finally, it is interesting to note that
\begin{mathletters}
\begin{eqnarray}
\left [ \nu^{^{ML}} (x^\mu) + {\cal C} \nu^{^{ML\,\dagger}} (x^\mu) \right
]/2 &=& \int {d^3 {\bf p} \over (2\pi)^3 } {1\over 2E_p} \sum_\eta \left
[\pmatrix{i\Theta \phi_{_L}^{\ast \, \eta} (p^\mu) \cr 0\cr} a_\eta
(p^\mu)  e^{-ip\cdot x} +\right.\nonumber \\
&+&\left.\pmatrix{0\cr
\phi_L^\eta (p^\mu)\cr } a_\eta^\dagger (p^\mu) e^{ip\cdot x} \right ]\,
,\\
\left [\nu^{^{ML}} (x^\mu) - {\cal C} \nu^{^{ML\,\dagger}} (x^\mu) \right
]/2 &=& \int {d^3 {\bf p} \over (2\pi)^3 } {1\over 2E_p} \sum_\eta \left
[\pmatrix{0\cr \phi_{_L}^\eta (p^\mu) \cr } a_\eta (p^\mu)  e^{-ip\cdot x}
+\right.\nonumber\\
&+&\left.\pmatrix{-i\Theta \phi_{_L}^{\ast\, \eta} (p^\mu)\cr 0
\cr } a_\eta^\dagger (p^\mu) e^{ip\cdot x} \right ]\, .  \end{eqnarray}
\end{mathletters}
thus naturally leading to the Ziino-Barut scheme of massive chiral
fields, ref.~\cite{Ziino}.

The conslusion is: it is still required a lot of work to make certain
conclusions about the relevance of the presented construct to describing
the physical world and to the present situation in the neutrino physics.
But, it is important that this construct is permitted by the requirements
of the extended Poincar\'e group symmetry; it is based on the very
viable postulates: in fact, after imposing the coniditions of
self/anti-self charge conjugacy we derived all consequences only on the
basis of the Wigner rules for transformations of left- and right- handed
2-spinors and on the relations between these spinors in the frame with
zero momentum.

Thus, as I was tought  in the Gorbachev's epoch:  ``everything is
permitted unless forbidden".

\end{itemize}

\end{document}